\documentstyle[11pt,newpasp,twoside]{article}
\def\edcomment#1{\iffalse\marginpar{\raggedright\sl#1\/}\else\relax\fi}
\reversemarginpar

\begin{document}
\title{Gravitational Lensing and Elliptical Galaxies}
\author{Daniel J.\ Mortlock}
\affil{Astrophysics Group, Cavendish Laboratory, Madingley Road,
        Cambridge CB3 0HE, United Kingdom}
\affil{Institute of Astronomy, Madingley Road, Cambridge CB3 0HA,
        United Kingdom}
\author{Rachel L.\ Webster}
\affil{School of Physics, University of Melbourne, Parkville, Vic 3052,
	Australia}


\section{Introduction}

The probability that high-redshift quasars are gravitationally-lensed
by intervening galaxies increases rapidly with the cosmological
constant, $\Omega_{\Lambda_0}$ (whilst being only weakly dependent
on the density parameter, $\Omega_{{\rm m}_0}$), and the 
low number of lenses observed implies that $\Omega_{\Lambda_0} \la 0.7$
(e.g.\ Kochanek 1996). One of many uncertainties has been the 
(small) core radii of elliptical galaxies, which, at least naively,
reduce their lensing cross-section. However, if ellipticals are
normalised relative to their observed line-of-sight velocity dispersion,
$\sigma_{||}$, then increasing the core radius must result in 
an increased mass normalisation (specified by the assymptotic 
velocity dispersion, $\sigma_\infty$). 

\section{Elliptical galaxies}

Elliptical galaxies are modelled as having
de Vaucouleurs (1948) surface brightness
profiles with (non-singular) isothermal mass distributions --
constant mass-to-light ratio models cannot fit both lensing 
observations and dynamics (Koch\-anek 1996). 
With a few other non-pathological assumptions about the dynamics,
solution of the Jeans equation is sufficient to calculate 
the line-of-sight dispersion as a function of projected radius, which can
then be integrated over the central regions of the galaxy, to 
link $\sigma_{||}$, $\sigma_\infty$, and the core radius, $r_{\rm c}$. 
Whilst there are some model uncertainties, this approach 
shows that $\sigma_\infty \simeq 1.1 \sigma_{||}$ for singular models,
and that $\sigma_\infty$ increases roughly linearly with $r_{\rm c}$
(Mortlock \& Webster 2000).

\section{Lensing probability}

With the choice of standard models for the galaxy population
and the (optical) quasar luminosity function,
the expected fraction of lenses in a magnitude-limited quasar
sample, $p_{\rm q}$, can be calculated. Under the incorrect
assumption that $\sigma_\infty = \sigma_{||}$,
$p_{\rm q}$ decreases with the scale core
radius of the galaxies, $r_{{\rm c}*}$, because, despite the increased
magnification bias, the reduction in cross-section dominates.
However the use of a self-consistent dynamical normalisation 
changes this dependence markedly, as shown in Fig.\ 1. 
Most importantly, the self-consistent
lensing probability never drops below the maximum 
probability for the incorrect models.

\begin{figure}
\includegraphics{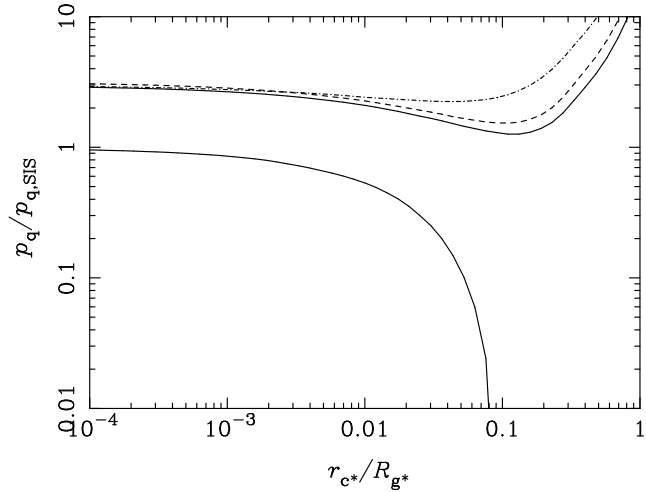}
\vspace{67mm}
\caption{The dependence of the lensing probability (relative to
the naive singular model) on the canonical
core radius of elliptical galaxies. With the identification 
$\sigma_\infty = \sigma_{||}$ the lower solid line is obtained; 
using a self-consistent dynamical normalisation results in the 
upper trio of lines, which show how the results depend on the 
cosmological model: 
$\Omega_{\rm m_0} = 1$ and $\Omega_{\Lambda_0} = 0$ (solid line);
$\Omega_{\rm m_0} = 0$ and $\Omega_{\Lambda_0} = 0$ (dashed line);
and
$\Omega_{\rm m_0} = 0$ and $\Omega_{\Lambda_0} = 1$ (dot-dashed line).} 
\label{figure:p_lens}
\end{figure}

\section{Conclusions}

The use of self-consistent dynamical models for elliptical galaxies
shows that, independent of their core radius, they are effective
lenses. The results imply that the current upper limits 
placed on $\Omega_{\Lambda_0}$ due to the low observed frequency of quasar
lenses are not weakened by any uncertainty in the core structure of 
ellipticals -- in fact the existing limits are probably more 
robust than previously believed.

\acknowledgments

DJM was supported by an Australian Postgraduate Award during the 
time this research was carried out.


\begin{references}

\reference de Vaucouleurs, G., 1948, Annalen d'Astrophysics, 11, 247

\reference Kochanek, C.\ S., 1996, \apj, 466, 638

\reference Mortlock, D.\ J., \& Webster, R.\ L., \mnras, in press

\end{references}
\end{document}